\documentclass{article}
\usepackage{graphicx}
\usepackage{emulateapj}

\slugcomment{\today}

\begin{document}

\def\etal{{\it et al.~}}
\def\eg{{\it e.g.,~}}
\def\ie{{\it i.e.,~}}

\title{The MHD Kelvin-Helmholtz Instability III:\\
       The Role of Sheared Magnetic Field in Planar Flows\altaffilmark{4}} 

\author{Hyunju Jeong\altaffilmark{1},
        Dongsu Ryu\altaffilmark{1},
        T. W. Jones\altaffilmark{2},
        and Adam Frank\altaffilmark{3}}

\altaffiltext{1}
{Department of Astronomy \& Space Science, Chungnam National University,
Daejeon 305-764, Korea: jhj@astro6.chungnam.ac.kr and
ryu@canopus.chungnam.ac.kr}
\altaffiltext{2}
{Department of Astronomy, University of Minnesota, Minneapolis, MN 55455:
twj@msi.umn.edu}
\altaffiltext{3}
{Department of Physics and Astronomy, University of Rochester, Rochester
NY 14627: afrank@alethea.pas.rochester.edu}
\altaffiltext{4}
{Accepted by the Astrophysical Journal}

\begin{abstract}

We have carried out simulations of the nonlinear evolution of 
the magnetohydrodynamic (MHD) Kelvin-Helmholtz (KH) instability for
compressible fluids in $2\frac{1}{2}$-dimensions, 
extending our previous work by Frank \etal (1996)
and Jones \etal (1997).
In the present work we have simulated flows in the $x-y$ plane in which
a ``sheared'' magnetic field of uniform strength ``smoothly'' rotates
across a thin velocity shear layer from the $z$ direction to the $x$
direction, aligned with the flow field.
The sonic Mach number of the velocity transition is unity.
Such flows containing a uniform field in the $x$ direction are linearly
stable if the magnetic field strength is great
enough that the Alfv\'enic Mach number, ${M_A} = {U_0}/{c_A} < 2$.
That limit does not apply directly to sheared magnetic fields, however,  since
the $z$ field component has almost no influence on the linear stability. 
Thus, if the magnetic shear layer is contained within the velocity
shear layer, the KH instability may still grow, even when the field strength
is quite large. So, we consider here 
a wide range of sheared field strengths covering Alfv\'enic Mach numbers,
${M_A} = 142.9$ to $2$. 

We focus on dynamical evolution of fluid features, kinetic energy 
dissipation, and mixing of the fluid between the two layers, considering
their dependence on magnetic field strength for this geometry. There
are a number of differences from our earlier simulations
with uniform magnetic fields in the $x-y$ plane. 
For the latter, simpler case we found a clear sequence of behaviors with
increasing field strength ranging from
nearly hydrodynamic flows in which the instability evolves to an almost steady
Cat's Eye vortex with enhanced dissipation, to flows in which the
magnetic field disrupts the Cat's Eye once it forms, finally, to flows that
evolve very little before field-line stretching stabilizes the
velocity shear layer. 
The introduction of magnetic shear can allow a Cat's Eye-like
vortex to form, even when the field is stronger than the nominal 
linear instability limit given above.
For strong fields that vortex is asymmetric with respect to the preliminary
shear layer, however, so the subsequent dissipation is enhanced over the
uniform field cases of comparable field strength.
In fact, so long as the magnetic field achieves some level of dynamical
importance during an eddy turnover time, the asymmetries introduced
through the magnetic shear will increase flow complexity, and, with that,
dissipation and mixing.
The degree of the fluid mixing between the two layers is 
strongly influenced by the magnetic field strength. 
Mixing of the fluid is most effective when the vortex is disrupted by 
magnetic tension during transient reconnection, through local chaotic
behavior that follows.

\end{abstract}

\keywords{instabilities -- methods: numerical -- MHD -- plasmas}

% \clearpage

\section {Introduction}

The Kelvin Helmholtz (KH) instability is commonly expected in boundary layers 
separating two fluids and should occur
frequently in both astrophysical and geophysical environments. 
The instability taps the free energy of the relative motion between two regions
separated by a shear layer, or ``vortex sheet''
and is often cited as a means to convert that directed flow energy into
turbulent energy (\eg Maslowe 1985).
Astrophysically, the instability is likely, for example, along jets generated 
in some astrophysical sources, such as active galactic nuclei
and young stellar objects (\eg Ferrari \etal 1981). 
There are also strongly sheared flows in the solar corona (\eg Kopp 1992).
Geophysically, it is expected on the earth's magnetopause separating
the magnetosphere from the solar wind (\eg Miura 1984).

The KH instability leads to momentum and energy transport, as well as
fluid mixing. If the fluid is magnetized, then the 
instability can locally amplify the magnetic field
temporarily by stretching it,
enhance dissipation by
reconnection, and lead to self organization between the magnetic and
flow fields. The linear evolution 
of the KH instability was thoroughly studied 
a long time ago. The analysis of the basic
magnetohydrodynamic (MHD) KH linear stability was 
summarized by Chandrasekhar (1961) and Miura \& Pritchett (1982) and
has been applied to numerous astrophysical 
situations (\eg Ferrari \etal 1981; Bodo \etal 1998). 
Chandrasekhar showed, for example, that for incompressible flows, a uniform
magnetic field aligned with the flow field will stabilize the shear layer if
the Alfv\'enic Mach number of the transition is small enough; in particular,
for a plasma of uniform density if $M_A = U_0/c_A < 2$, where $U_0$ 
is the velocity change across the shear layer.
There is now a considerable literature on the nonlinear evolution of the
hydrodynamical (HD) KH instability (\eg Corcos \& Sherman 1984; Maslowe 1985)
thanks to the development of both robust 
algorithms to solve fluid equations and fast computers to execute numerical 
simulations. However, the literature on the nonlinear behavior of the
MHD KH instability  is much more limited.
That is because MHD physics is more complicated and because accurate,
robust numerical 
algorithms, especially for compressible flows, are relatively new.
Since a great many astrophysical applications surely involve magnetized
fluids, it is important to come to a full understanding of the MHD version of
the problem.

In the past several years initial progress has been made,
especially in two-dimensional MHD flows involving initially
uniform magnetic fields having a component aligned with the flow field 
(\eg Miura 1984; Wu 1986; Malagoli \etal 1996; Frank \etal 1996 [Paper I];
Jones \etal 1997 [Paper II]).
The KH problem is, of course, a study of boundary evolution. 
An interesting and physically very relevant variation on this problem 
adds a current sheet to the vortex sheet. 
Then, the direction and/or the magnitude of the magnetic field will
change across the shear layer.
Dahlburg \etal (1997) recently presented a nice discussion of the linear
properties of this problem for non-ideal, incompressible MHD
when $M_A \sim 1$,
and also examined numerically some 
aspects of the early nonlinear evolution of current-vortex sheets.
They pointed out for the (strong field) parameters
considered that the character
of the instability changes depending on the relative widths of the
current and vortex sheets. Let those be measured by $a_b$ and $a_v$,
respectively. When the vortex sheet was thicker than the
current sheet (actually when $(a_b/a_v) M_A^{(2/3)} < 1$),
they found the instability to be magnetically dominated
and that its character resembled resistive, tearing modes; \ie ``sausage
modes''. 
In the opposite regime, on the other
hand the instability evolved in ways that were qualitatively similar
to the KH instability, but with dynamically driven magnetic
reconnection on ``ideal flow timescales'', independent of the resistivity.
Recently, Keppens \etal (1999) explored numerically the situation in which the 
magnetic field reverses direction discontinuously within the shear layer.
They found that this magnetic field configuration actually enhances the
linear KH instability rate, because it increases the role of tearing 
mode reconnection, which accelerates plasma and helps drive circulation.
Miura (1987) and Keller \& Lysak (1999) reported some results
from numerical simulations with sheared magnetic fields inside a velocity
shear layer; that is, with magnetic fields whose direction rotates
relative to the flow plane. Those papers considered 
specific cases designed to probe issues associated with convection
in the earth's magnetosphere. Both found significant influences from
the geometry of the magnetic field.
Galinsky and Sonnerup (1994) described three-dimensional simulations
also with sheared magnetic fields inside a velocity shear layer.
Their simulations followed the early nonlinear evolution of three-dimensional
current-vortex tubes, but were limited by low resolution.

The work reported here provides an extension through compressible MHD
simulations of the results given by Dahlburg \etal (1997).
More explicitly it extends our earlier studies of Papers I and II
in this direction.
To summarize the latter two studies:
Paper I examined the two-dimensional nonlinear evolution 
of the MHD KH instability on periodic sections of unstable sheared flows of a 
uniform-density, unit Mach number plasma with a uniform 
magnetic field parallel to the flow direction.
That paper considered two different magnetic field strengths corresponding to
$M_A = 2.5$, which is only slightly weaker than required for linear
stability, and a weaker field case, $M_A = 5$.
It emphasized that the stronger field case became nonlinearly
stable after only a modest growth, resulting in a stable, laminar flow.
The weaker field case, however, developed the classical ``Kelvin's Cat's Eye''
vortex expected in the HD case. That vortex, which is a stable
structure in two-dimensional HD flows,
was soon disrupted in the $M_A = 5$ case by
magnetic tension during reconnection, so that this flow also became
nearly laminar and effectively stable, because the shear layer was much
broadened by its evolution.
Subsequently, Paper II considered in $2\frac{1}{2}$-dimensions a wider
range of magnetic field strengths for the same plasma flows, 
and allowed the still-uniform
initial field to project out of the computational plane, but
in a direction still parallel to the shear plane,
over a full range of angles,
$\theta=\arccos({{\vec B}\cdot{\hat x}}/|{\vec B}|)$.
With planar symmetry in $2\frac{1}{2}$-dimensions,
the $B_z$ component out of the plane interacts only
through its contribution to the total pressure. So, Paper II emphasized
for the initial configurations studied there that the 
$2\frac{1}{2}$-dimensional nonlinear evolution of
the MHD KH instability was entirely determined by the Alfv\'enic Mach 
number associated with the field projected onto the computational plane; 
namely, $M_{Ax} = (U_0/c_A)\cos{\theta}$.
Those simulations also showed how the magnetic field wrapped into the
Cat's Eye vortex would disrupt that structure if the magnetic tension
generated by field stretching became comparable to the centripetal force
within the vortex. That result is equivalent to expecting
disruption when the Alfv\'enic Mach
number on the perimeter of the vortex becomes reduced to values near
unity. 
This makes sense, because then the Maxwell stresses in the vortex
are comparable to the Reynolds stresses.
Since magnetic fields are ``expelled'' from a vortex by reconnection on 
the timescale of one turnover time (Weiss 1966) and
in two-dimensions the perimeter
field is amplified roughly by an order of magnitude during a turnover,
Paper II concluded that vortex disruption should
occur roughly when  the initial
$M_{Ax} \lesssim 20$.
For our sonic Mach 1 flows, this corresponds to plasma $\beta_0 \lesssim 480$,
which has a field strength range often seen as too weak to be 
dynamically important.
After the Cat's Eye was disrupted in those cases the flow settled into an
almost laminar form containing a broadened, but hotter shear layer that
was stable to perturbations smaller than the length of the computational
box.
For weaker initial fields the role of the magnetic field became primarily
one of enhanced dissipation through magnetic reconnection,
although the transition
between these two last behaviors is not sharp. 

Thus, for uniform initial magnetic fields one can define four distinct
regimes describing the role of the magnetic field
on the evolution of the $2\frac{1}{2}$-dimensional
KH instability. In descending field-strength order they are: 1) ``linearly
stabilized'', 2) ``nonlinearly stabilized'', 3) ``disruptive'' of the vortical
structures formed by the HD instability, and 4) ``dissipative'' in the sense
that the field enhances dissipation over the HD rate.

To extend our understanding to a broader range of field configurations,
we now study $2\frac{1}{2}$-dimensional cases in which the magnetic field 
rotates from the $z$ direction to the $x$ direction
within the shear layer.
Such field rotations are commonly
called ``magnetic field shear'' (\eg Biskamp 1994).
Magnetic shear will likely occur in a number of KH unstable
astrophysical environments. One example is the propagation of
astrophysical jets.  The current consensus hold that jets in both
Young Stellar Objects and and Active Galactic Nuclei are generated
via some form of magneto-centrifugal mechanism (Ouyed \& Pudritz 1997)
which produces a jet beam with a helical topology.  The environment
however will likely have fields that differ substantially from
from the tight helix in the beam. Thus the KH unstable shear layer
at the beam-environment interface will also be a region where the field
rotates, as well.

Consistent with the behaviors identified by Dahlburg \etal (1997) for
related field structures
we find that this modification strongly alters the simple 
patterns we found for
uniform fields and adds new insights to the roles played by the magnetic
fields. 
In our new simulations the magnetic shear layer is fully contained within
the velocity shear layer; that is, the vortex sheet and the current sheet
are the same width.
Our current sheet is oblique to the flow field, however, so that one
one side of the velocity shear magnetic tension is absent in 2 1/2-dimensions.
In this situation {\it the shear layer always remains KH unstable.}
This is true even for field strengths that would have
stabilized flows were the field uniform.
At the beginning of each simulation
the interactions between the flow and the magnetic
field resemble those described for uniform fields
in the one half of the space
where the magnetic field lies within the flow plane. 
In the other half plane, however,
the initial magnetic field interacts only
through its pressure, so has a negligible
influence. This situation breaks the symmetry inherent in our
earlier simulations. The resulting nonlinear evolution of the KH instability
can be much more complex, and this has considerable influence on the
dissipation rate and the mixing that takes place between fluids initially
on opposite sides of the shear layer.
Those issues will be the focus of this paper.

The plan of the paper is as follows. In section II we summarize 
numerical methods and initial condition used in the simulations. 
In section III we compare and analyze the evolutionary results.
In section IV we contain summary and conclusion.

\section {Numerical methods and Initial Conditions}

Motion of a conducting, compressible fluid carrying a 
magnetic field must satisfy the MHD equations
consisting of the Maxwell's equations 
and the equations of gas dynamics, extended to include the influences of
the Maxwell stresses. 
In the MHD limit the displacement current and the separation between ions 
and electrons are neglected. The ideal, compressible MHD equations, 
where the effects of viscosity, electrical resistivity and thermal 
conductivity are neglected, can be written in conservative form as    
\begin{equation}
\frac{\partial \rho}{\partial t}+{\nabla}\cdot\left(\rho{\vec v}\right) = 0,
\label{continuity equation}
\end{equation}
\begin{equation}
\frac{\partial(\rho{\vec v})}{\partial t} + \nabla_j\cdot\left(\rho
{\vec v}v_j-{\vec B}B_j\right)+\nabla\left(p+{1\over2}B^2\right) = 0,
\label{momentum equation}
\end{equation}
\begin{equation}
\frac{\partial E}{\partial t} + \nabla\cdot\left[\left(E+p+{1\over2}B^2
\right){\vec v}-\left({\vec v}\cdot{\vec B}\right){\vec B}\right] = 0,
\label{energy equation}
\end{equation}
\begin{equation}
\frac{\partial \vec B}{\partial t} + \nabla_j\cdot\left({\vec B}v_j
-{\vec v}B_j\right) = 0,
\label{equation of magnetic field evolution}
\end{equation}
where gas pressure is given by
\begin{equation}
p=\left(\gamma-1\right)\left(E-{1\over2}\rho v^2-{1\over2}B^2\right).
\end{equation}
The units are such that magnetic pressure is $p_B = 1/2 B^2$ and the
Alfv\'en speed is simply $c_A = B/\sqrt{\rho}$.

In our study these equations were solved
using a multi-dimensional MHD TVD code 
using a Strange-type directional splitting
(Ryu \& Jones 1995; Ryu \etal 1995). 
It is based on an explicit, second-order Eulerian finite-difference scheme
called the Total Variation Diminishing (TVD) scheme 
which is a second-order-accurate extension of 
the Roe-type upwind scheme. This version of the code contains an fft-based 
routine that uses ``flux cleaning'' to maintain the 
${\nabla}\cdot {B} = 0$ condition at each time step within machine accuracy.

We simulated physical variables
(${\rho}$, ${\rho}{v_x}$, ${\rho}{v_y}$, ${\rho}{v_z}$, ${B_x}$,
${B_y}$, ${B_z}$, $E$) in the $x-y$ plane 
on a computation domain ${{x}=[0,L_x]}$ and ${{y}=[0,L_y]}$  
with ${{L_x}={L_y}=L=1}$. Here we used  periodic conditions 
on the ${x}$ boundaries and reflecting conditions on the ${y}$ 
boundaries. The mass, kinetic energy and Poynting fluxes in the $y$
direction all vanish  at $y=0$  and $y=L$.
The total magnetic flux through the box remains constant
throughout the simulations.
Such boundary conditions were used initially by Miura (1984) 
and subsequently in Papers I and II.

As in the two previous papers we simulated an initial background
flow of uniform density, ${\rho=1}$, gas pressure, ${p=0.6}$, and
adiabatic index, $\gamma=5/3$, so that the initial sound speed is 
${c_s}= \sqrt{{\gamma}p/{\rho}}=1$. We considered a
hyperbolic tangent initial velocity profile that 
establishes nearly uniform flow
except for a thin transition layer in the mid-plane of the simulations. 
Explicitly, we set
\begin{equation}
{v_0}(y)\hat x = - {{U_0}\over 2} \tanh\left({{y-{L/2}}\over a}\right)
{\hat x}.
\end{equation}
Here ${U_0}$ is the velocity difference between the two layers and was
set to unity, so that the sonic Mach number of the transition is also
unity; \ie $M_s = U_0/c_s = 1$.
This equation describes motion of fluid flowing to the right in the 
lower part (${0 \le y \le{L/2}}$) of the two layers and to the left 
in the upper part (${{L/2} \le y \le {L}}$).
To reduce un-wanted interactions with the $y$ boundaries, and to assure an
initially smooth transition on a discrete grid, $a$ should 
be chosen to satisfy $ {h < {a} << L}$ where $h$ is the size of 
the computational grid cells.
Here we considered $a={L/25}$, and $h \approx (1/10)a$.
These choices were evaluated in Paper I.

We began with a magnetic field of uniform strength, but one
that rotates (\ie is sheared) smoothly within the transition layer 
($L/2-a < y < L/2+a$) from the $z$ direction 
on the bottom of the grid to the $x$ direction on the top; namely,
\begin{equation}
{B_x}={B_0},~~{B_z}={0}~~~
{\rm for}~~~{{L\over 2}+a} < y < L,
\end{equation} 
\begin{equation}                            
{B_x}={B_0}\sin\left({\pi\over 2} {{y-{L/2}+a}\over{2a}}\right)~~~
{\rm for}~~~{{L\over 2} - a} \le y \le {{L\over 2} + a},
\end{equation}
\begin{equation}
{B_z}={B_0}\cos\left({\pi\over 2} {{y-{L/2}+a}\over{2a}}\right)~~~
{\rm for}~~~{{L\over 2} - a} \le y \le {{L\over 2} + a},
\end{equation}
\begin{equation}
{B_z}={B_0},~~{B_x}={0}~~~
{\rm for}~~~0 < y < { {L\over 2}-a}.
\end{equation}
This construction keeps $|B|$ constant and gives nominally equal widths to the 
shear layer and the current sheet. 
The magnetic field shear is, however, sharply
bounded by $L/2 - a \le y \le L/2 + a$, so that the vortex sheet is
effectively slightly broader. As mentioned in the introduction, the
analysis of Dahlburg \etal (1997) leads to the expectation that perturbations
on this initial set up will be unstable to the KH instability modes,
even when $M_A < 1$.

The other relevant MHD parameters are defined as $\beta_0 = {p}/{p_b}$, 
which measures relative 
gas and magnetic pressures, and the Alfv\'enic Mach number of the
shear transition, $M_A = U_0/c_A$. These are related through the
relation $M^2_A = (\gamma/2)M^2_s \beta_0$.
We have considered a wide range of these
parameters as listed in Table 1.
A random perturbation of small amplitude was added to the velocity to initiate
the instability.

All the simulations reported here were
carried out on a grid with $256\times 256$ cells.
Papers I and II included convergence studies for similar flows using
the same code. There we found with the resolution used in the
present study all major flow properties that were formed
in equivalent, higher resolution simulations, and that such global
measures as energy dissipation and magnetic energy evolution were
at least qualitatively similar. The key was having sufficient
resolution that there was an essentially non-dissipative
range of flow scales inside the Cat's Eye vortex. That is, it was
necessary that the effective kinetic and magnetic Reynolds numbers
are large for the Cat's Eye. The effective Reynolds numbers scale
as the square of the number of cells for this code (Ryu \etal 1995), so
that is readily achieved in the current simulations.

An additional comment is appropriate on the numerical methods.
The code used for these studies nominally treats the flows as ``ideal'',
or non-dissipative. Dissipation does take place, of course, through numerical
truncation and diffusion at the grid-cell level; \ie 
primarily within the smallest resolved
structures. The consistency of this approach
with non-ideal HD flows of high Reynolds number has been convincingly 
demonstrated using turbulence simulations for
conservative methods analogous to those employed here 
(\eg Porter \& Woodward 1994; Sytine \etal 1999).
While that comparison has not yet been accomplished for MHD flows,
there are a number of results that support consistency for ideal MHD
codes when the dissipation scales are small, as well. These include
the apparent ``convergence'' in general flow and magnetic field
patterns and global energy evolution seen in our own simulations 
mentioned above, as well as MHD turbulence studies 
(\eg Mac Low \etal 1998; Stone \etal 1998).
In addition, others among our quasi-ideal MHD simulations develop
structures that behave like those generally ascribed to resistive reconnection,
such as unstable current sheet tearing-modes (\eg Miniati \etal 1999) when
the effective Lundquist number is large and stable Parker-Sweet
current sheets  when that parameter is not large
enough to be tearing-mode unstable (Gregori \etal 1999).

\section {Results}

\subsection {Structure Evolution}

Table 1 includes a descriptor in each case to indicate,
for reference, the evolution expected
for equivalent simulations using a uniform
magnetic field orientation instead of the sheared field. For example, the 
strongest field case with $M_A = 2$ would be marginally stable if the field 
were uniform.
Based on Papers I and II the uniform-field $M_A = 2.5$ and $3.3$ cases
would be ``nonlinearly stable'', the $M_A = 5,\,10$ and $14.3$ cases would be 
``disruptive'', while the two weakest field cases ($M_A = 50,\,142.9$)
would be ``dissipative''.  Recall that the disruptive cases were those
whose initially weak fields were amplified through stretching around 
the vortex to the point that the vortex was destroyed.

As previously stated, the MHD KH evolution beginning from a sheared field
in the velocity shear layer can be very different
from the comparable uniform magnetic 
field case.
{\it Because the magnetic field cannot stabilize flows in those regions where
it is perpendicular to the flow plane, 
the linear KH instability can always develop there,
independent of the strength of the field}.
For the flows investigated here
the HD linear growth time $\tau_g \sim 1.5 \lambda$ in simulation units,
where $\lambda$ is the wavelength of the perturbation.
On a slightly longer timescale one or more ``large'' vortices
will generally form.
But if magnetic tension on the aligned-field side of 
the shear layer is dynamically
significant, the vortex or vortices will not be symmetric across the
shear layer and the subsequent nonlinear 
evolution of the full flow field can 
become complex through interactions between
the two regions.
In this section we will outline briefly the behavior patterns we observe in
our sheared magnetic field simulations as a function of field strength, 
beginning with the very weak field cases.

In our simulations, cases with extremely weak magnetic fields 
(${M_A}=142.9$ and ${M_A} = 50$) produce a relatively symmetric, 
stable Cat's Eye vortex essentially the same as in HD flows. As we found
in Paper II, it spins indefinitely on the timescales considered here.
Similarly to the uniform field cases considered in Paper II, 
the magnetic flux initially in
the region where the field is aligned with the flow 
($L/2+a < y < L$) is stretched around 
the vortex, increasing magnetic energy. However, in these cases the
magnetic field amplification by stretching during one vortex turnover is not
adequate to reduce the local Alfv\'enic Mach number on the perimeter of
the vortex to values near unity.
Recalling that the vortex turnover time is also the timescale to
generate magnetic topologies unstable to the tearing mode instability and
driven reconnection (see, \eg Papers I and II), the magnetic field
in these cases is not able to disrupt the quasi-HD nature of the flow.
As for the analogous uniform-field cases we studied before, 
the inclusion of a very weak magnetic field
mostly enhances kinetic energy dissipation via the dissipation that comes
as a consequence of magnetic reconnection. There is little new understanding
that is added from those cases.

In cases with field values in the ``disruptive'' regime for uniform fields
(${M_A} = 14.3$, ${M_A} = 10$ and ${M_A} = 5$) the magnetic field is too
weak still to inhibit initial formation of vortices in either half of
the flow field, but sufficiently strong to destroy those vortices
after the external field pulled into the vortex perimeter
becomes stretched by vortex rotation.
In the sheared magnetic field flows the Maxwell stresses on the Cat's Eye are
asymmetric, since there is no flux pulled into the vortex from one side.
As one considers stronger fields this dynamical asymmetry becomes more obvious.
In fact, in the sheared field $M_A = 5$ case the Cat's Eye never 
forms fully, whereas it
did in the uniform field simulation for the same field strength (Paper I).
Evolution of the $M_A = 14.3$ case is illustrative of the weaker
sheared field evolution and is shown in Fig. 1.
A single Cat's Eye develops by the first time shown in the figure ($t = 7$),
but is clearly in the processes of disruption by the third time ($t = 10$). 
This early evolution closely parallels the results for the 
$M_A = 5$ uniform field case discussed in Paper I. But, already by $t = 10$
the symmetry of the flow in Fig. 1 is obviously broken. 
The subsequent evolution
qualitatively resembles the later ``weak field'' 
evolution shown in Paper I, but with a
very interesting twist. The analogous flow in Paper I developed a ``secondary''
vortex in the mid-plane of the flow. This structure was created by the
sudden release of magnetic tension during reconnection associated with
disruption of the Cat's Eye. The magnetic flux in the interior of the secondary
vortex was isolated from that of the exterior flow, creating a ``flux island''.
However, the secondary vortex was ``spun down''
by magnetic tension around its perimeter, 
since the vortex was embedded in magnetic
flux crossing the computational box. Thus, there was a net torque
provided by the magnetic tension around the vortex. 
Magnetic flux in the interior of
the secondary vortex was annihilated through reconnection as the
secondary vortex was dissipated.
In the uniform field simulation of Paper I this stage was part of a
relaxation of the entire flow to a broadened, nearly
laminar shear layer, and the magnetic and flow fields were nearly perfectly
aligned, that is, they ``self-organize''. 
Paper I referred to this as a ``quasi-steady relaxed'' state.

In the present cases there are multiple ``secondary  vortices'' generated
at various locations within the grid as the primary, Cat's Eye
vortex is disrupted, as one might 
expect from a random initial perturbation. Each secondary vortex also
holds a magnetic island. Secondary vortices that become entrained
within the part of the flow where the shear is greatest
contain the greatest enstrophy so rotate very obviously.
Here, enstrophy is defined as $(\nabla\times v)^2\sim (\frac{1}{2} \omega^2)$ 
where $\omega$ is the rotational frequency of a vortex.
The secondary vortices tend to merge, as one
expects in two-dimensional flow,
since they all have the same sign in vorticity. But, 
in the $M_A = 14.3$ case shown in Fig. 1
a vortex pair still remain at the end of the simulation.
Once again, these structures hold closed magnetic field loops, 
or flux islands. 
During vortex merging the magnetic flux islands 
also merge through reconnection.
However, unlike the
uniform field case discussed in Paper I, these structures and the flux 
within them then tend to be stable to the end of the simulation.
The reason for this difference is that in the present case the secondary
structures that survive are all in the lower part of the flows where
there is almost no magnetic flux crossing the box. So, their external
interactions are almost entirely those of quasi-ideal HD.
By contrast, the upper portions of these flows, 
where $\vec B$ was initially in the
flow plane, fairly quickly become smooth and laminar. Almost all of
the open field lines within the $x-y$ plane are now concentrated into
this region and well aligned
with the velocity field. Thus, this portion of the flow resembles the
quasi-steady relaxed state of Paper I.
Also, as for the flow studied in Paper I, the aligned
magnetic flux becomes concentrated near the velocity shear layer.
In the present case, however, the shear layer is displaced upwards to
some degree;
that is, towards the region where the magnetic field was initially
aligned to the computational plane. There is just a little compression
in these flows, so this shift represents a net exchange of momentum
between the two layers as a consequence of vortex disruption, rather than
a squeezing of the fluid in one half of the plane.

On the whole we find that the disruptive behavior by a
weak field acts sooner in the sheared magnetic field cases than
was the situation with a uniform field of the same strength. This point was
already mentioned with regard to comparing flows with $M_A = 5$, and
seems generally to be a direct consequence 
of the symmetry breaking in the sheared
field cases. Symmetry breaking causes deformations in the Cat's Eye, which, in 
turn, enhance the degree of field-line stretching around its perimeter.
In the $M_A = 5$ sheared field case, for example, a secondary vortex
rolls up at the end of the Cat's Eye during its formation, pulling
magnetic field with it. That field quickly becomes disruptive, 
the main Cat's Eye breaks up, and the central shear layer
relaxes into a quasi-steady flow. For this particular case, early disruption of
the vortical flow means a significant reduction in the kinetic energy
dissipation within the flow compared to the equivalent uniform field
case. On the other hand the asymmetry should be responsible for
the net momentum exchange mentioned in the previous paragraph.

The consequences of beginning with a sheared, strong field are somewhat
different, however. By strong we mean that a uniform field in the 
flow plane would lead either to linear stability or nonlinear 
stability of the shear layer, as described earlier.
In sheared field cases (${M_A} = 3.3$, ${M_A} = 2.5$ and ${M_A} = 2$),
that is still the situation on the side of the flow
where the magnetic field initially lies within the flow plane. On the
other side, however, the magnetic field initially has a negligible
influence, only adding an isotropic pressure. Here, just as for an HD
flow, corrugations of the shear layer 
grow because of the Bernoulli effect. Meanwhile, 
aligned magnetic field across the shear layer
can be stretched out by the growing
corrugation, stiffening and, if it is strong, 
maintaining a smooth flow where the
field is in the plane. Shear becomes concentrated along the
resulting inflection in the flow boundary, causing a vortex to be
shed into the quasi-HD portion of the flow. This secondary vortex formation
is the same process that was mentioned previously for
the somewhat weaker field case with $M_A = 5$. It is illustrated
clearly for three strong field cases in Fig. 2. Once the secondary vortex is
shed in these cases, the shear layer stabilizes, and the
vortex becomes embedded in the quasi-HD portion of the flow.
The vortex contains a magnetic island similar to those discussed
earlier.  Once again, this structure also seems to be stable and
long lived in our simulations.

\subsection {Energy Evolution: Dissipation}

In Papers I and II we pointed out the role played by a magnetic field
in determining the amount of kinetic energy dissipated during evolution
of a KH unstable shear layer and also examined the evolution of
magnetic energy during flow development. Here we revisit those issues briefly
in order to  see how the differing dynamical influence of the magnetic
shear affects energy evolution within the flow. Recall that some
amount of kinetic energy must be dissipated during the formation of
Kelvin's Cat's Eye, since the kinetic energy of the latter flow
pattern is less than in the initial flow. For the initial conditions
used in our simulations the kinetic energy reduction expected in a
two-dimensional HD simulation is about 7\%,
and since the initial kinetic energy is
about 10\% of the initial thermal energy, the analogous increase in thermal
energy is roughly 0.7\% of the total.
This transition is irreversible, of course.
Generally, we find the dissipation to be enhanced when a magnetic field
is added. The greatest energy dissipation comes from disruption of the
Cat's Eye, since portions of the flow become chaotic, and
those chaotic motions are quickly dissipated. When the magnetic field is
strong enough to prevent initial vortex roll-up (nonlinearly stable cases),
viscous dissipation accompanies flow and it generally exceeds the dissipation
associated with the HD Cat's Eye flow. Even a very weak magnetic field
that has no obvious direct dynamical consequences
enhances energy dissipation, since driven magnetic reconnection
and magnetic energy annihilation in the perimeter of the Cat's Eye
is also irreversible. 

Since the dynamical influence of the sheared magnetic field is
different from that of an equivalent uniform field aligned in the
flow plane, we also should expect at least some quantitative
differences from the results reported in Papers I and II.
Figs. 3, 4 and 5 allow us to explore this.
Figs. 3 and 4 show the time evolution of the 
thermal energy ($E_t$), kinetic energy ($E_k$), 
magnetic energy ($E_b$) for the sheared field simulations.
Each quantity is normalized by its initial value.
The figures also show the minimum value of the plasma 
$\beta $ parameter in the computation as a function of time. 
The simulations are grouped into those with fields
weak enough to allow fully formed
Cat's Eye structures (Fig. 3) and those with fields strong enough
to prevent that formation, as discussed above (Fig. 4).

Energy dissipation associated with Cat's Eye formation is clearly evident
between $t \sim 6$ and $t \sim 8$ in Fig. 3 
from the evolution of the thermal and kinetic energies.  The changes in those
two quantities are about the values just cited for Cat's Eye generation.
The subsequent energy evolution varies dramatically with $M_A$, however.
For the very weak field cases with $M_A = 142.9$ and $50$,
the subsequent dissipation is small, but larger than for a purely HD flow. 
The dissipation rate is greater for the $M_A=50$ case than for
the $M_A=142.9$ case,
since there is more magnetic energy being generated and dissipated, as
discussed in Paper II. For the cases with $M_A=14.3$ and $M_A=10$,
disruption of the Cat's Eye between $t \sim 10 - 20$ leads to much greater
subsequent kinetic energy dissipation. 
In those two cases the kinetic energy reduction
is also significantly greater than in the cases with comparable
field strength in our uniform field simulations.
For $M_A=14.3$, for example, the kinetic
energy at the end of the analogous uniform field simulation in Paper II
was reduced
by about 50\%, whereas here the final kinetic energy is less than 25\%
of the initial value. That added dissipation comes from the more
extensive and complex flow structures entraining magnetic flux
formed in the sheared magnetic field evolution.

Fig. 4 shows that the energy dissipation pattern is quite different for the 
flows with stronger magnetic fields. There is a period of moderately
large dissipation as the initial corrugations of the shear layer 
grow to saturation and secondary
vortices are shed and dissipated. Afterwards, however,
the dissipation rate is only
modestly greater than for quasi-HD flows such as the $M_A = 142.9$ case, 
since the flow pattern for these cases is laminar in the aligned field region 
and uniform
except for isolated vortices in the transverse field region (see Fig. 2).
In the $M_A = 3.3$ case the interface between
these two regions is 
less regular than the other three cases, so that flow remains a little 
more dissipative to the end.
The dissipation dependence on $M_A$ in all our sheared field simulations
is summarized in Fig. 5, which shows the amount of the kinetic energy left 
at the end of simulations ($t=40$).

Figs. 3 and 4 also show us the evolution of magnetic energy during the
simulations. There are evident
differences between the two groupings. In particular
there is a relatively small fractional increase in magnetic energy for
the stronger field cases. That makes sense, of course, because
in those cases magnetic field is never stretched around a vortex.
The peak enhancement to the magnetic energy in those cases
ranged from less than 6\% for the $M_A = 2$ case to
$\sim25\%$ for the $M_A = 3.3$ and $M_A = 5$
cases.  Those peaks correspond to periods of maximum stretching
by growth of the initial corrugation at early times, $t \sim 7-10$.
In the $M_A = 5$ case, the peak corresponds to an extended interaction
between the magnetic field in the shear layer
with the secondary vortex, where the magnetic field {\it is} wrapped up.
As the flows relax in these cases, the magnetic energy returns to values close 
to, but slightly below the initial values.
Since the magnetic flux through the grid is conserved, this lower
final magnetic energy must come from the fact that flux in $B_z$, initially
entirely in the lower half of the box has now encroached into the
upper half (see Figs. 7, 9), while flux in $B_x$ has expanded
from the top half of the box into the lower half. 

By contrast, the peak magnetic energy enhancement in the
weaker field simulations ranges from factors of $\sim3.5$ for the
$M_A = 10$ case to $\sim7.8$ for the $M_A = 14.3$ case.
The peak values here correspond to formation of the Cat's Eye near $t = 8$
and subsequent episodes of magnetic field stretching.
(See Fig. 1 for the $M_A = 14.3$ case.)
In all of these cases the final magnetic energy is at least as large as
the initial value, in contrast to the situation described for the
stronger field cases. The reason for excess residual magnetic energy is
the formation of coherent ``flux tube'' like structures similar
to those described in Paper I.
In the $M_A = 14.3$ case the final magnetic energy is about
twice its initial value. For the others in this group the differences are
much smaller, however.

\subsection {Mixing Across the Boundary Layer}

We have emphasized how the addition of magnetic shear across the
unstable flow boundary adds complexity to the nonlinear flow that
follows from the instability. So far we have seen how that increases
the dissipation of kinetic energy from the flow over the HD solution
and over the equivalent MHD solution with a uniform magnetic
field. This added complexity may also influence mixing between
fluids initially in the top and bottom layers. Figs. 6, 7 and 8 give us
some insight into that process. They illustrate how magnetic flux 
perpendicular to the flow plane has been transported from the lower
part of the computational domain into the top part. Recall that initially
$B_z = 0$ for $y > L/2 + a = 0.54$ (Eqs. 7-10), 
so that $\int^{top}B_z dxdy/\int^{bottom}B_z dxdy = 0.03$ in Fig 6.
Note, also, that the total $B_z$ flux through the plane is constant. Since
this component is frozen into the fluid very well and the fluid
is almost incompressible during these simulations, the evolution of
this ratio in Fig 6 offers a simple way to examine approximately
how fluid from the lower part of the flow region penetrates into the
upper flow.
In the strong field cases, generally
less than $\sim15\%$ of the perpendicular magnetic flux penetrates the
upper flow. That penetration is mostly due to the spreading of the
shear layer, which is illustrated in Fig. 9 for these cases by the form of
$\int v_x(x,y) dx/L$ at the end of simulations.
Fig. 7 also helps clarify that point by showing the
maximum upward penetration of the $B_z$ component during each
simulation. There is a clear separation in this quantity between the
strong field cases and weak field cases. In the strong field cases the
maximum penetration of $B_z$ is generally around $y = 0.7$, which
corresponds about to the location of the top of the shear layer at the end
of these simulations, as shown in Fig. 9.

By contrast, Fig. 6 shows that the weaker field cases transport between 
$\sim 1/4$ and $\sim 1/3$ of the $B_z$ flux into
the top part of the computational
domain. In these cases the transport happens primarily during the period
when the instability is in transition from
linear to nonlinear evolution. That point is illustrated in more detail
for the weak field, disruptive case with $M_A = 14.3$ in Fig. 8.
There we see at various times $\int B_z(x,y) dx/L$.
By $t = 20$, $B_z$ has achieved almost its maximum penetration in $y$.
Cat's Eye disruption for this case takes place
between roughly $t = 10$ and $t = 20$ (see Fig. 1). By the end of this
simulation perpendicular magnetic flux is roughly uniformly distributed
below $y \sim 0.7$. As Fig. 1 illustrates, that value of $y$ also
corresponds to the lower bound for most of the $B_x$ flux in the relaxed
flow. In fact, most of the $B_x$ flux is concentrated into a
``flux tube'' just above $y = 0.7$, accounting for the fact that the
final magnetic energy for this case is significantly greater than the
initial magnetic energy. On the other hand, the center of the velocity
shear is still close the center of the grid, where the two secondary
vortices are located (visible in Fig. 1).
Thus, since the $B_z$ flux is effectively frozen into its initial fluid,
we can see that the disruption of the Cat's Eye has resulted in considerable
exchange and mixing of fluids between the top and bottom portions of the
computational domain. Through the Maxwell stresses generated in the 
magnetic field lying within the flow plane, that exchange has also
included exchange of momentum, since at the end the velocity distribution
is still not far from symmetric.

\section {Summary and Conclusion}

We have performed numerical simulations of the nonlinear 
evolution of MHD KH instability in $2\frac{1}{2}$-dimensions. 
A compressible MHD code has been used.
However, with an initially trans-sonic shear layer ($M_s=1$),
flows stay only slightly compressible ($\left<\delta\rho/\rho\right> <<1$),
and compressibility plays only minor roles (see \S 3.1).
We have considered initial configurations in
which the magnetic field rotates across the velocity shear layer
from being aligned with the flow on one side to being perpendicular
to the flow plane on the other; that is, we have considered ``sheared''
magnetic fields. The velocity and magnetic shears have similar
widths. In the top portion of the flow,
where the magnetic field is aligned, magnetic tension can inhibit
the instability or at least modify its evolution, as
we and others described in earlier papers. 
In the bottom portion of the flow
the magnetic field has negligible initial influence, at least in
$2\frac{1}{2}$-dimensions. We have simulated flows with a
wide range of magnetic field strengths.
Our objective was to understand how this
symmetry break would modify our earlier findings and hence to gain
additional insights to the general properties of the MHD KH instability.

When the magnetic field is extremely weak, the flows are virtually
hydrodynamic in character except for added dissipation associated with
magnetic reconnection on small scales. So the field geometry does not appear 
to be very 
important. We should emphasize once again, however, that
one should be careful in choosing to label magnetic fields as ``very
weak'', because our current findings support our earlier conclusion
that an initially weak field embedded in a complex flow can have
crucial dynamical consequences. In our simulations
of trans-sonic shear layers, we find that magnetic fields (whether
sheared or not) can disrupt the HD nature of the flow even when the
initial $\beta_0 = p/p_b \sim$ several hundred. Actually, the more meaningful
parameter is the Alfv\'enic Mach number, $M_A$, 
since that provides a comparison
between the Maxwell and Reynolds stresses in a complex flow. 
In the KH instability, particularly, the free energy derives directly from
directed flow energy, not isotropic pressure.
We find for
all of our $2\frac{1}{2}$-dimensional simulations that flows beginning
roughly with $M_A\lesssim20$ will be strongly modified by the magnetic fields
during the nonlinear flow evolution. These characteristic values are
commonly seen as outside the ranges where MHD is the necessary language,
but that is evidently not the case.

For MHD flows initiated with $M_A\lesssim20$,
we find some interesting differences
between the flows beginning with uniform and sheared magnetic fields.
Magnetic tension will in those cases become significant by stretching
of field lines out of the portion of the initial flow containing
the aligned field. It may even inhibit development of the
instability in that space. But, for sheared fields flow
on the side with the perpendicular field begins
to evolve almost hydrodynamically, so the instability always
evolves into a nonlinear regime, whereas for uniform fields
cases with $M_A < 2$ are stable for the flow properties we are considering.
Further, for sheared fields the resulting nonlinear structures are
highly asymmetric and tend to become chaotic. That, in turn, tends to
increase kinetic energy dissipation and mixing of plasma when
compared to flows starting with uniform fields. 

In previously simulated flows beginning 
with uniform magnetic fields that became
dynamically important, the end result of prolonged evolution was always
a quasi-laminar flow with a broadened velocity shear layer. In the 
sheared magnetic field simulations presented here, however, 
that description is not
complete. It does adequately characterize the eventual flows on the
side of the shear layer containing the aligned magnetic field. Also, on the
other side, the flows do become smooth to some degree and appear
stable. However, there are typically embedded vortices within
the flow region that does not have significant aligned magnetic flux.
Those vortices contain
magnetic flux islands with long lifetimes. We expect they will
eventually dissipate, but apparently only on timescales much longer
than our simulations consider.

In conclusion, we find that applying sheared magnetic fields across
a KH unstable boundary of comparable width still leads to the same qualitative
weak-field behaviors (dissipative and disruptive) that we identified
in Paper II.
For stronger fields the instability is not inhibited by the presence of
the field, however. The sheared fields also enhance the complexity within
the flow and generally increase the amount of kinetic energy dissipation
and fluid mixing that comes from the instability.

Boundary layers that include sheared flows and sheared magnetic fields are 
probably astrophysically common, so it is of some consequence to note
how the vortex and current sheets behave together. The best known
examples are the earth's magnetopause (\eg Russell 1990) and 
coronal helmut streamers (\eg Dahlburg \& Karpen 1995) both of  which
may appear to be magnetically dominated. Other relevant situations
might include ``interstellar MHD bullets'' (\eg Jones \etal 1996; Miniati
\etal 1999; Gregori \etal 1999), where initially weak magnetic fields 
become stretched around the cloud to form a magnetic shield. If there is
a tangential discontinuity on the cloud surface, then it could
resemble the situation we discuss. The magnetosheaths of comets may
also include current and vortex sheets (\eg Yi \etal 1996). 
Our present work 
supports and augments the findings of Keppens \etal (1999) that 
the instabilities developing on apparently magnetic dominated boundaries
containing velocity and
magnetic shear may still retain some 
of their hydrodynamical character. On the other hand, this work also
reinforces our previous results showing that even a relatively weak field
embedded in an unstable vortex sheet tends to smooth the nonlinear flow.
Detailed comparisons with astrophysical objects should be done with
caution, however, since these numerical studies are rather idealized, and not
fully 3D. The extension to 3D will be discussed elsewhere 
(Jones \etal 1999; Ryu \etal 1999).

\acknowledgments

The work by HJ and DR was supported in part by KOSEF through
grant and 981-0203-011-2.
The work by TWJ was supported in part by the NSF through grants
AST93-18959, INT95-11654, AST96-19438 and AST96-16964, by NASA grant 
NAG5-5055 and by the University of Minnesota Supercomputing Institute.
AF was supported in part by NSF Grant AST-0978765 and the University of 
Rochester's Laboratory for Laser Energetics.
We thank anonymous referee for clarifying comments.

% \clearpage

% \clearpage

\begin{figure}
% \epsfxsize=16truecm 
% \centerline{\epsfbox{fig1.ps}}
\figcaption[Fig1]{Evolution of gas density (top panels) and 
magnetic field lines (bottom panels) for the case with $M_A=14.3$. 
High tones correspond to low values.
The times shown are $t = 7,\,9,\,10,\,15,\,17.5,\,40$.}
\end{figure} 
 
\begin{figure}
% \epsfxsize=16truecm
% \centerline{\epsfbox{fig2.ps}}
\figcaption[Fig2]{Evolution of gas density for the cases with
(top to bottom) $M_A=3.3$, $M_A=2.5$, $M_A=2$. 
High tones correspond to low values.
Times shown are $t = 6.5,\,15.5,\,40$ for $M_A=3.3$;
$t = 7.5,\, 15.5,\, 40$ for $M_A=2.5$;
$t = 8.5,\, 15.5,\, 40$ for $M_A=2$.}
\end{figure}

\begin{figure}
% \epsfxsize=16truecm
% \centerline{\epsfbox{fig3.ps}}
\figcaption[Fig3]{Time evolution for the weaker magnetic field
cases of the thermal energy ($E_t$), kinetic energy ($E_k$), and
magnetic energy ($E_b$) normalized to their initial values.
The minimum value of the plasma $\beta~(={p_g}/{p_b})$ parameter is
also shown.
Shown cases have $M_A = 142.9,\,50,\,14.3,\,10$.
A Cat's Eye vortex forms in each of these flows.}
\end{figure}

\begin{figure}
% \epsfxsize=16truecm
% \centerline{\epsfbox{fig4.ps}}
\figcaption[Fig4]{Same as Fig. 3, except for the stronger field
cases in which no Cat's Eye vortex becomes fully formed.
Shown cases have $M_A = 5,\,3.3,\,2.5,\,2$.}
\end{figure}

\begin{figure}
\epsfxsize=16truecm 
\centerline{\epsfbox{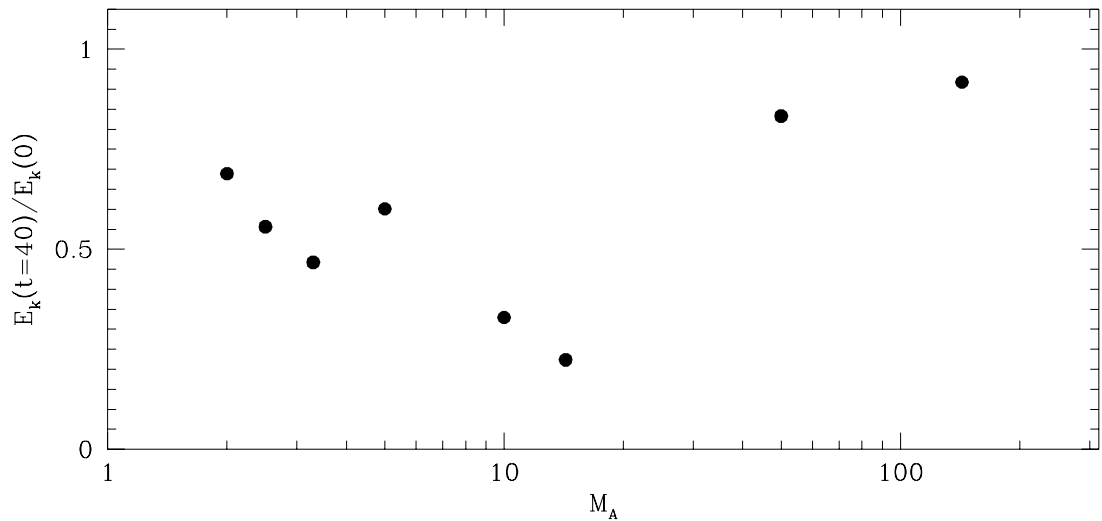}}
\figcaption[Fig5]{The kinetic energy left at the end of simulations
($t=40$). It is normalized to the initial value.
This provides a measure of kinetic energy dissipation.}
\end{figure}

\begin{figure}
\epsfxsize=16truecm
\centerline{\epsfbox{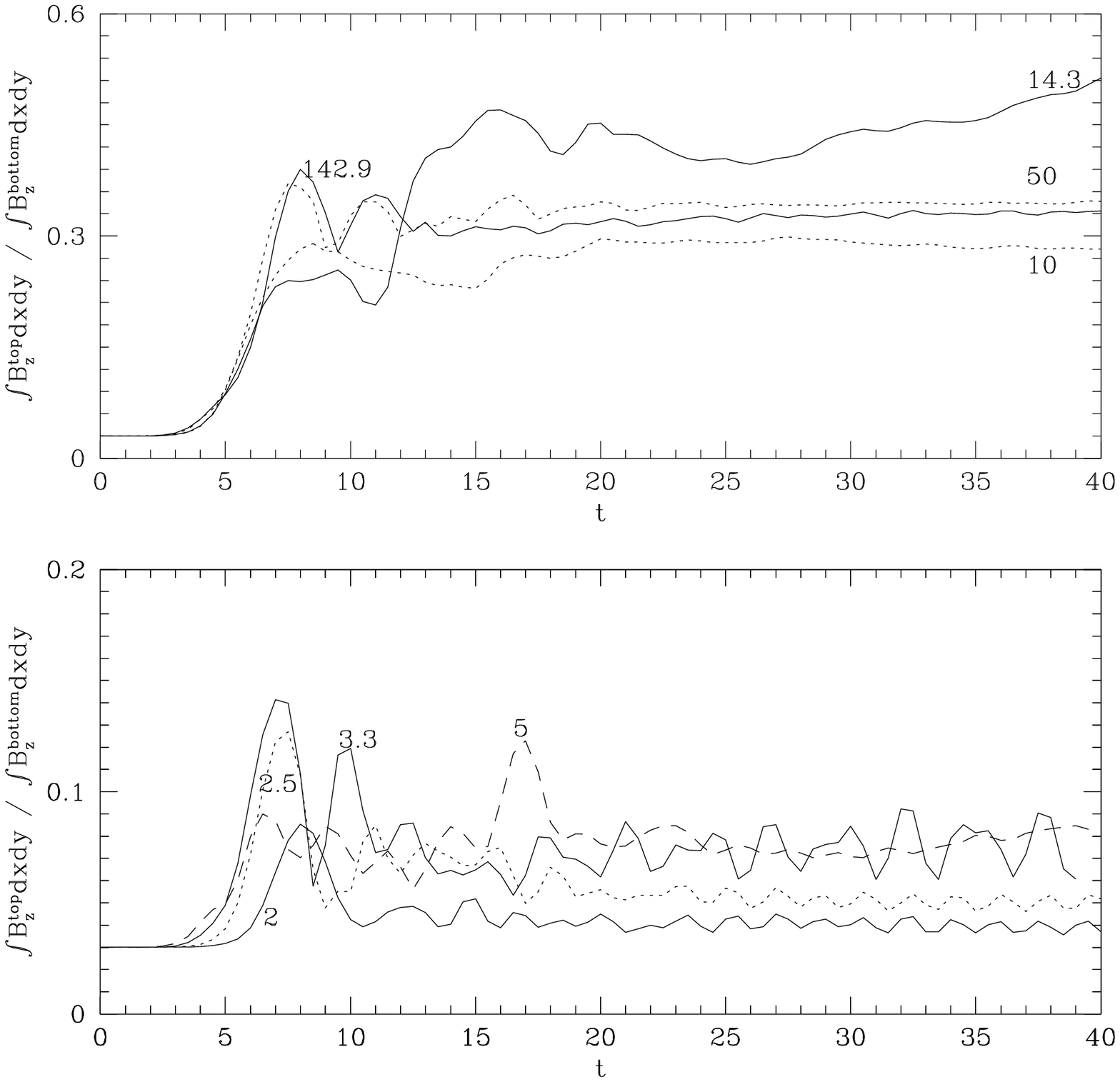}}
\figcaption[Fig6]{Time evolution of the ratio of the mean $B_z$ 
in the top region, ${L/2} < y < L$, over that in the bottom region,
$0 < y < {L/2}$.
Curves are labeled with $M_A$.}
\end{figure}

\begin{figure}
\epsfxsize=16truecm
\centerline{\epsfbox{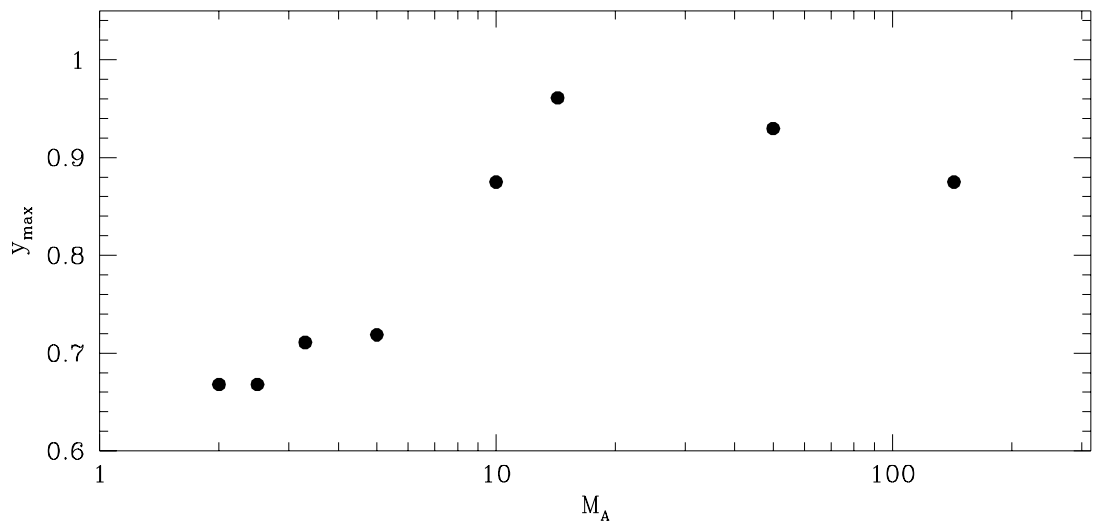}}
\figcaption[Fig7]{Maximum extent of non-zero $B_z$ in the
$y$ direction during the simulations.}
\end{figure}

\begin{figure}
\epsfxsize=16truecm
\centerline{\epsfbox{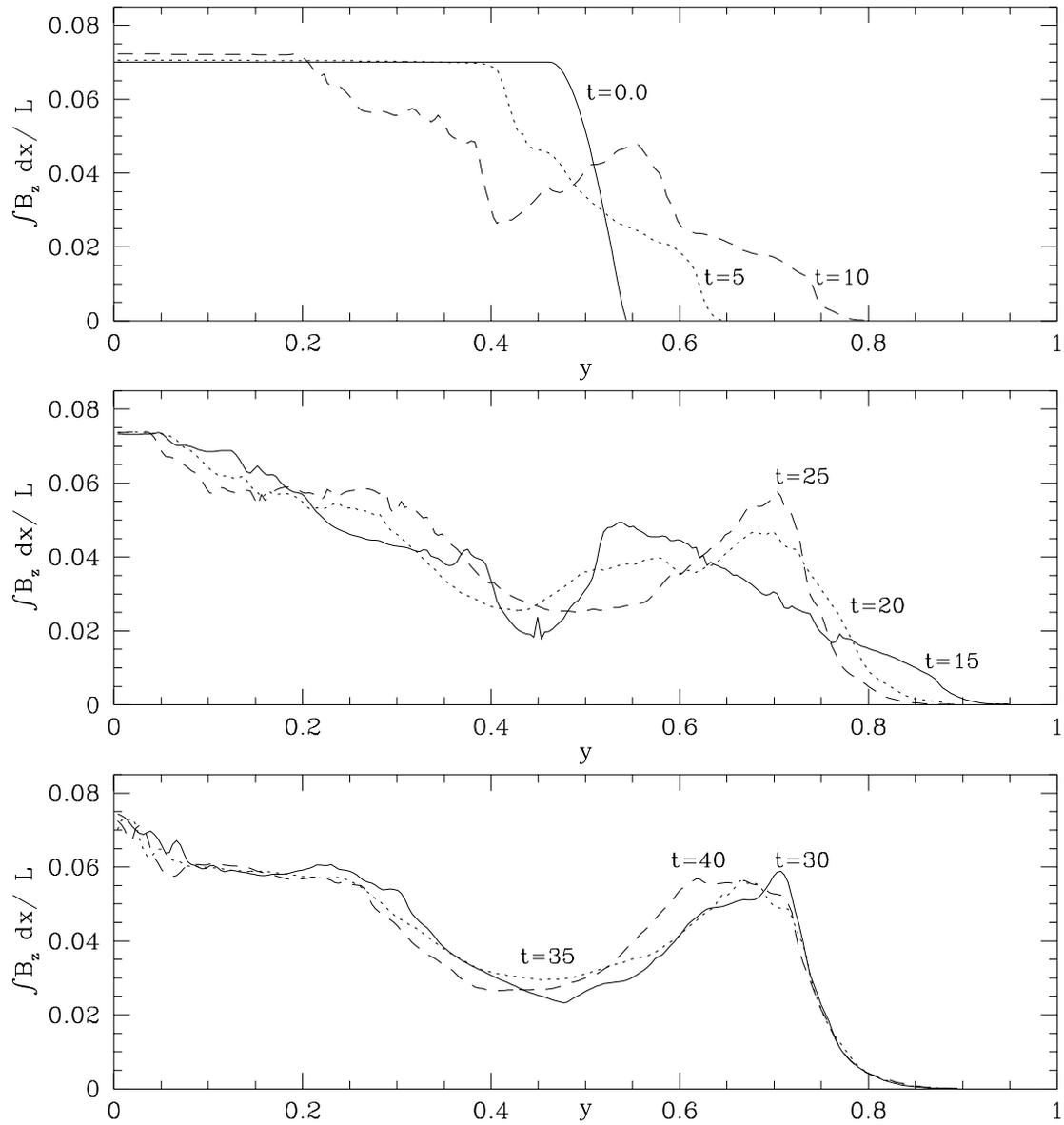}}
\figcaption[Fig8]{Evolution of the $x$ average value of $B_z$
for the case with $M_A = 14.3$.}
\end{figure}
 
\begin{figure}
\epsfxsize=16truecm
\centerline{\epsfbox{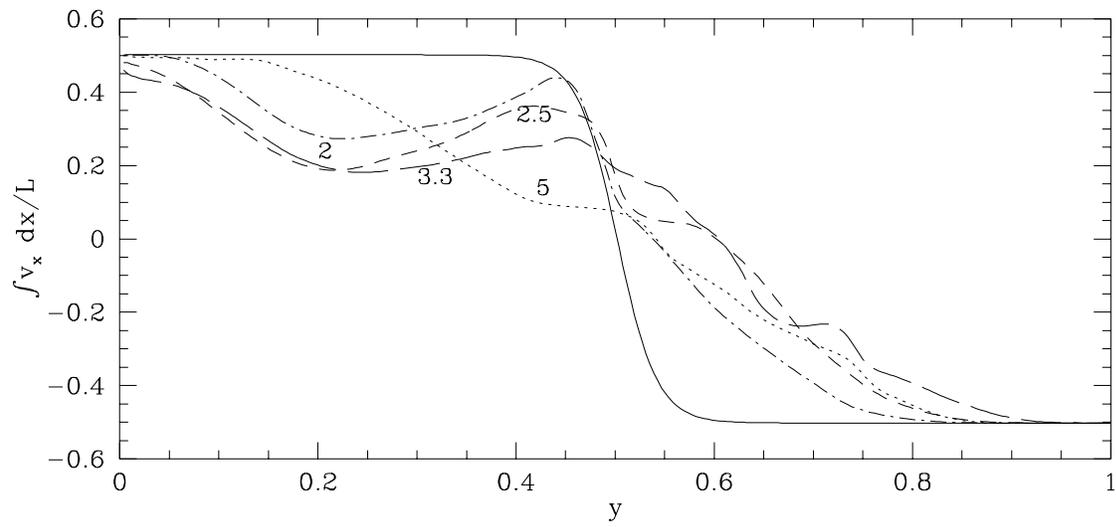}}
\figcaption[Fig9]{The mean velocity profile as a function
of $y$ at $t = 40$ for the stronger field cases.
Curves are labeled with $M_A$.}
\end{figure}

% \clearpage

\begin{deluxetable}{ccccccc}
\footnotesize
\tablecaption{Initial Parameters of Simulations.\tablenotemark{a}}
\tablehead{\colhead{$B_0$} & \colhead{$M_A (=U_0/c_A)$} &
\colhead{$\beta_0 = p/p_b$} &
\colhead{Uniform Field Evolution\tablenotemark{b}} &
\colhead{Sheared Field Evolution\tablenotemark{c}} }
\startdata
$0.007$ & $142.9$ & $2.4 \times 10^4$ & dissipative & dissipative \nl
$0.02$ & $50$ & $3000$ & dissipative & dissipative \nl
$0.07$ & $14.3$ & $245$ & disruptive & disruptive \nl
$0.1$ & $10$ & $120$ & disruptive & disruptive \nl
$0.2$ & $5$ & $30$ & disruptive & asymmetric \nl
$0.3$ & $3.3$ & $13.3$ &  nonlinearly stable & asymmetric \nl
$0.4$ & $2.5$ & $7.5$ &  nonlinearly stable & asymmetric \nl
$0.5$ & $2$ & $4.8$ &  marginally stable & asymmetric \nl
\enddata
\tablenotetext{a}{All simulations have been carried out with
$\gamma = 5/3$, $\rho = 1$, $U_0 = 1$, $M_s=U_0/c_s=1$, $L=1$ 
and $a=L/25$, using a $256\times 256$ grid up to $t=40$.}
\tablenotetext{b}{Evolution character expected for a uniform,
non-sheared magnetic field (see Papers I and II).}
\tablenotetext{c}{Evolution character observed for a non-uniform,
sheared magnetic field (see text).}
\end{deluxetable}

\end{document}